\documentclass[preprint,12pt]{aastex}

\newcommand{\be}{\begin{equation}}
\newcommand{\ee}{\end{equation}}
\newcommand{\nn}{\mbox{} \nonumber \\ \mbox{} }
\newcommand{\ba}{\begin{eqnarray}}
\newcommand{\ea}{\end{eqnarray}}
\newcommand{\om}{\omega}

\newcommand\etal{\textit{et al.\ }}
\newcommand\eg{\textit{e.g.\ }}
\newcommand\cf{\textit{cf.\ }}

\newcommand{\Bf}{{magnetic field\,}} 
\newcommand{\Bfs}{{magnetic field\,}}

\newcommand\cxo {{\it Chandra}}

\newcommand{\tfe}{1E~1048.1--5937} 
\newcommand{\tfn}{1E~2259.1+586}

\newcommand{\soe}{1RXS~1708--4009}

\newcommand{\ett}{XTE~J1810--197}

\newcommand{\efs}{RX~J1856.5--3754}

\def\lapp{\ifmmode\stackrel{<}{_{\sim}}\else$\stackrel{<}{_{\sim}}$\fi}
\def\gapp{\ifmmode\stackrel{>}{_{\sim}}\else$\stackrel{>}{_{\sim}}$\fi}
\begin{document}

\title{Resonant Cyclotron Scattering and  Comptonization in Neutron Star Magnetospheres} 
\author{Maxim Lyutikov}
\affil{ Department of Physics and Astronomy, University of British Columbia,
6224 Agricultural Road, Vancouver, BC, V6T 1Z1, Canada \\ and \\
Department of Physics and  Astronomy, University of Rochester,
Bausch and  Lomb Hall,
P.O. Box 270171,
600 Wilson Boulevard,
Rochester, NY 14627-0171 }
\author{Fotis P. Gavriil}
\affil{Physics Department, McGill University, 3600 rue University
Montreal, QC,  H3A 2T8, Canada
}


\date{Received   / Accepted  }

\begin{abstract}
Resonant cyclotron scattering of the surface radiation in the
magnetospheres of neutron stars may considerably modify the emergent
spectra and impede efforts to constraint neutron star properties.  Resonant
cyclotron scattering by a non-relativistic
warm plasma in an {\it inhomogeneous} magnetic field has a number of unusual
characteristics: (i) in the limit of high resonant optical depth, the
cyclotron resonant layer is {\it half opaque}, in sharp contrast to the case
of non-resonant scattering. (ii) The {\it transmitted flux is on average
Compton up-scattered} by $\sim 1+ 2 \beta_{th}$, where $\beta_{th}$ is the
typical thermal velocity in units of the velocity of light; the
reflected flux has on average the initial frequency.  (iii) For both
the transmitted and reflected fluxes the dispersion of intensity {\it
decreases} with increasing optical depth. (iv) The emergent spectrum
is appreciably non-Plankian while narrow spectral features produced at
the surface may be  erased.  

We derive  semi-analytically modification of the surface Plankian
emission due to multiple scattering between the resonant layers and
apply the model to anomalous X-ray pulsar \tfe.  Our simple model fits
just as well as the ``canonical'' magnetar spectra model of a
blackbody plus power-law.
\end{abstract}

\section{Introduction}

The magnetospheres of neutron stars are filled with plasma, with the
minimum magnetospheric plasma density being on the order of the
Goldreich-Julian density $n_{GJ} = {\bf B} \cdot {\bf \Omega}/ 2 \pi e
c$ ($ {\bf B}$ is \Bf, ${\bf \Omega}$ is angular velocity, $e$ is the
elementary charge and $c$ is the velocity of light),
\cite{goldreich69}.  The magnetospheric plasma may efficiently scatter
and absorb surface emission through resonant cyclotron scattering.  If
the plasma density in the magnetosphere equals the minimum
Goldreich-Julian density, then resonant scattering is not important in the
optical-X-ray range \citep[][Eq.~(\ref{tau01})]{Mikh82}. On the other
hand, under certain circumstances the real plasma density in the
magnetosphere may exceed this minimum value.  In rotationally powered
pulsars pair excess on closed field lines can be due to pair creation
by high energy photons \cite[\eg][]{wrhz98}.  In magnetars
\citep[\eg][]{wt04}, large scale currents flowing in the magnetosphere
result in particle densities much larger than the Goldreich-Julian
density \citep{tlk02}.  In accreting sources (\eg neutron stars in
X-ray binaries or isolated neutron stars accreting from the
interstellar medium, ISM) the external plasma may penetrate inside the
magnetosphere due to the development of instabilities in the boundary
layer \citep{gpl77,al80}.  Finally, the presence of a dense
magnetospheric plasma is well established in the case of the binary
pulsar PSR J0737-3039B, where the multiplicity factor (the ratio of
plasma density to Goldreich-Julian density) was inferred to be $\sim
10^5$ \citep{lt05}.

The presence of a relatively tenuous plasma may lead to large
magnetospheric optical depths for scattering and absorption since in
soft X-rays the resonant cyclotron cross-section is five to six orders
of magnitude larger than the Thomson cross section,
Eq.~(\ref{sigmadigma}). Due to the large resonant cross-section,
the amount of electron-ion plasma that must be
suspended in the magnetosphere to produce an optical depth of the
order of unity is tiny by astrophysical standards, of the order of
$10^9$ grams total, Eq.~(\ref{N}).  Both the emission patterns and
spectral properties of the surface radiation are then modified in the
magnetosphere, as photons are scattered and their energies are
Doppler-shifted in each scatter.

In this paper we describe salient properties of resonant scattering in
neutron star magnetospheres and investigate semi-analytically the
influence of scattering on emergent spectra in the non-relativistic
limit, when the plasma temperatures are smaller than $mc^2/k_B$ (where
$m$ is the electron mass and $k_B$ is Boltzmann's constant).  We
employ methods of radiative transfer that were developed for
line-driven winds \citep{chan45,sobo47} and were previously applied to
cyclotron scattering by \cite{Zhel96}.  We adopt a simple
one-dimension model for photon propagation (called the
Schwarzschild-Schuster method) which allows a clear description of the
underlying physics. 
More involved Monte-Carlo simulations which  take into account
the full three-dimensional structure of the magnetosphere as well as
relativistic effects are underway.

The majority of prior work on resonant cyclotron scattering considered
either the atmosphere or the accretion column of neutron stars
\citep{WS80,nag81,LMW90,derm91}, in which case the thickness of the
plasma column $H$ (centimeters or meters) was assumed to be much
smaller than the thickness of the resonant layer (see Section
\ref{layer}) $\Delta r \sim \beta_{th} r$ ($r$ is the distance from the
center of the neutron star and $ \beta_{th}$ is the thermal velocity in terms of
the velocity of light).  In this case the transfer of resonant
radiation occurs in a quasi-homogeneous magnetic field.  Higher up in
the magnetosphere $H$ becomes larger than $\Delta r$, so that the
effects of the \Bf's inhomogeneity become important.  In this case
properties of resonant scattering are different from the case of a
homogeneous \Bf and are somewhat counterintuitive. For example,
\cite{Zhel96} has pointed out that in a strongly inhomogeneous \Bf in
the high optical depth limit resonant layers are semi-opaque.
Following up on this work, we investigate spectral properties of the
transmitted radiation.  We show that if a considerably dense, warm
plasma is present in the magnetosphere (so that both the resonant
cyclotron optical depth and the frequency Doppler shifts during
scattering are non-negligible), then (i) the transmitted spectra are
modified, so that {\it thermal surface emission has a resulting
non-Plankian spectrum}; this effect is most pronounced at optical
depths $\sim$ few, while for very large optical depths the spectrum
remains nearly Plankian; (ii) {\it transmitted radiation is preferentially
Compton up-scattered} in frequency, so that the real surface
temperature may be smaller than the one inferred from spectral fits;
(iii) surface spectral features may be erased; (iv) scattering in the
optical and X-ray bands occurs in different regimes due to different
polarization properties of normal modes (circularly polarized in
optical and linearly polarized in X-rays). 

 We assume that plasma is tenuous, so that typical frequencies are
much larger than the plasma frequency, and the typical wavelength is much
larger than the skin depth. The refractive indexes are then close to unity
and Debye screening is not important in calculating the scattering
probabilities.  Collisional processes may also be neglected for the
densities and temperatures involved.
We also assume that initially radiation is not polarized; this is a considerable
simplification since neutron star surface emission is expected to be polarized, 
especially at magnetar field strengths \citep{ho-lai,ozel}. 

\section{Resonant Scatter in the Magnetosphere}
\label{layer}
 \subsection{Resonant Layers}

If the magnetosphere of a neutron star is populated by warm
(non-relativistic) plasma, quasi-thermal radiation coming from the
surface of the neutron star may experience strong scattering at the
cyclotron resonance, located for a given photon energy at a radius
\begin{equation} 
r \sim r_{NS} \left( { \hbar e B_{NS} \over \epsilon
m c} \right)^{1/3} \sim 8 r_{NS} b^{1/3} \left( {\epsilon \over 1 {\rm
keV}} \right)^{- 1/3}, 
\label{rres}
\end{equation} 
where $ r_{NS}\sim 10^6 $ cm is
the neutron star radius, $ B_{NS} $ is the surface magnetic field, $b=
B_{NS}/B_{cr}$, $B_{cr}= 4\times 10^{13}$ G is the critical magnetic
field,  $\hbar$ is Plank constant divided by $2\pi$,
$m$ is the electron mass, $ \epsilon $ is the energy of a photon.  The
 importance of resonant scattering may be characterized by the
 resonant optical depth \citep{Mikh82,Zhel96} 
\begin{eqnarray}
 && \tau_{res}= \int
 \sigma_{res} n dl = { \pi^2 e^2 n r \over m c \om_B} \left| {\partial \ln
 \omega_B \over \partial \ln r} \right|^{-1} (1+ \cos ^2 \alpha') =
 \tau_0 {( 1+ \cos ^2 \alpha') } \nn && \tau_0 = { \pi^2 e^2 n r \over
 3 m c \om_B},
\label{tau0}
\end{eqnarray}
 where $\alpha'$ is the angle between the incoming photon and the
local magnetic field, $\om_B = eB/mc$ is cyclotron frequency,
$\sigma_{res}$ is a resonant cross-section
\begin{equation} 
\sigma_{res} = {\sigma_T \over 4} { (1+ \cos ^2 \alpha') \omega^2
\over ( \omega - \omega_B)^2 + \Gamma^2/4},
\label{sigmares}
\end{equation} 
$\Gamma = 4 e^2 \om_B^2 / 3 m_e c^3$ is the natural width of the first
cyclotron line, $n$ is the plasma density and we assumed dipolar
fields with $ \left| {\partial \ln \omega_B / \partial \ln r}
\right|=3$.  The meaning of optical depth $\tau_0$ is that only
$e^{-\tau_0}$ fraction of photons coming from direction $\alpha'$
passes through the resonant layer without scattering.  This is {\it
different} from  the relative number of photons that passes through the resonant
layer after experiencing multiple scattering in the layer.

In the spectral range of interest absorbed photons are almost
immediately re-emitted since the synchrotron transition times $\sim
1/\Gamma$ are shorter than the dynamical time $r/c$ for
\begin{equation} r< \left(
\frac{4 B_{NS}^2 e^4 r_{NS}^6}{3 c^6 m^3} \right)^{1/5} \sim 7.6 \times
10^8 b^{2/5} \, {\rm cm}
\end{equation} 
Or, in terms of energy 
\begin{equation} 
\epsilon >
\left( \frac{27 c^{13} \hbar^5 m^4}{64 B_{NS} e^7 r_{NS}^3} \right)^{1/5}
=10^{-3} b^{-1/5} {\rm eV}. 
\end{equation} 
Only microwaves and longer waves suffer true absorption; all the
higher energy radiation is scattered.

The resonant optical depth (\ref{tau0}) greatly exceeds the
non-resonant optical depth for Thomson scattering $\tau_{T} \sim R n
\sigma_T$,
\begin{equation} 
{ \tau_{res} \over \tau_{T} } \sim { \pi \over 8} 
{ c \over r_e \omega_B} \sim 10^5  
\left( \frac{1\ \mathrm{keV}}{\hbar \omega_B} \right) 10^5
\label{sigmadigma}
\end{equation} 
where  $r_e = e^2
/mc^2$ is the classical radius of an electron.  Estimating the total
 Thus, the presence of even a tenuous plasma
in the magnetosphere may provide a substantial optical depth to
resonant scattering.  
The corresponding optical depth is 
\begin{equation} 
\tau_{res}\sim 
\frac{1}{3}\left( \frac{n}{n_{GJ}}\right)\left( \frac{r}{R_{lc}}\right)
\label{tau01}
\end{equation} 
so that minimal magnetospheric density is not sufficient to produce
considerable absorption, except near the light cylinder.
In order to produce an optical depth of the order of unity the
required density is 
\begin{equation} 
n \sim {3 B \over  \pi^2 e r} \sim {3 b \over 
\pi^2} { mc \over \hbar} {mc^2 \over e^2} {r_{NS}^3\over r^4} 
\end{equation} 
 Estimating the total
volume of the scattering material as $V \sim 4\pi r^3/3$, the total
number of scatterers required is only 
\begin{equation} 
N \sim { b \over \pi r_e \lambda_C } {r_{NS}^3 \over r} \leq 10^{33}
b.
\label{N}
\end{equation} 
If the plasma is electron-ion, then this corresponds to a mass of only
$\sim 2 \times 10^{9} b$ g. 

\subsection{Resonant Scattering in an Inhomogeneous Magnetic Field}

magnetic field \footnote{The magnetic field can vary in direction and
magnitude.  For a non-relativistic plasma the variation in magnitude
is more important than the variation in direction, except in a narrow
region near $\alpha' \sim \pi/2$, \citep{Zhel96}} the spatial resonance
width is 
\begin{equation} 
\Delta r \sim {\beta_{th} \over \partial_r \ln \omega_B | \cos \alpha|} =
\beta_{th} L_B \sim \beta_{T} r /3
\end{equation} 
where $L_B = - 1/ \partial_r \ln \omega_B$ and the last equality
assumes a dipole magnetic field.


Depending on whether the thickness of the scattering layer $H$ is
larger or smaller than the resonant width $\Delta r=\beta_{th} L_B$, the
radiation transfer occurs in two different regimes: a weakly
inhomogeneous field, $ \beta_{th} L_B \gg H$ (this is relevant to
radiation transfer in photosphere and polar column), and a strongly
inhomogeneous field $\beta_{th} L_B \ll H$, relevant for radiation
transfer in the magnetosphere \citep{Zhel96}.  In a weakly
inhomogeneous field a resonant photon that has been scattered once and
whose frequency has been modified due to the Doppler effect, has a final
frequency that typically lies outside of the resonant layer, so that
the photon is likely to leave the system.  For a continuum spectrum
falling on the resonant layer this leads to formation of a narrow
absorption line.  In a strongly inhomogeneous field, $\Delta r \ll H$,
a scattered photon typically remains in resonance and can experience
multiple scattering.  If a warm plasma, $\beta_{th} \leq 1$, fills a
considerable fraction of the magnetosphere, the resonant width is
smaller that the thickness of the plasma column, $H\sim r$, by a
factor of $\beta_{th}/3$, so that resonant scattering occurs in the  limit
of a strongly inhomogeneous field.  For relativistically hot plasma,
$\beta_{th} \sim 1$, the width of the resonance becomes comparable to the
inhomogeneity scale and the approach developed in this paper is not
valid anymore. Resonant scattering in the relativistic limit has been
studied in the  context of early models of GRBs and pulsar magnetospheres
\cite[\eg][]{LMW90,Brainerd92,sturner95,Lyub00,gontier}

\subsection{Polarization in resonance}
\label{polariz}

In neutron star magnetospheres both plasma and quantum effects
influence the propagation and wave absorption through polarization of
normal modes.  Depending on the polarization of normal modes, there are
two possible regimes of cyclotron radiation transfer. For weak \Bfs
and/or low frequencies, plasma effects dominate over quantum
effects. In this case waves propagating along the \Bf are circularly
polarized (if plasma is electron-ion) and in the case of warm plasma,
$\beta_{th} \leq 1$, radiation transfer occurs on the extraordinary wave,
while cross-sections for scattering of the ordinary wave is smaller by
$\beta_{th}^2$.  Alternatively, in a strong \Bfs and/or for high
frequencies, wave polarization is determined by quantum vacuum
effects.  In this case waves are linearly polarized (for parallel
propagation) and both modes are strongly scattered.

In resonance, waves remain linearly polarized (dominated by quantum
contributions) if the following condition is satisfied \citep{Zhel96}
\begin{equation} 
{\om_p^2 \over \om_B^2 \beta_{th}} \ll {1 \over 45 \pi} {e^2 \over \hbar
c} b^2.
\end{equation} 
Assuming that the optical depth is of the order of unity, $\om_p^2 \sim
\om_B c/r$, dipole geometry and $\beta_{th} =0.3$
 we can find the photon energy above which polarization is
determined by quantum vacuum corrections
\begin{equation}
\epsilon_c = 3^{3/4} \left( {\pi^3 c^{19} \hbar ^5 m^7 \tau^3 \over
e^7 r_{NS}^3 \beta_{th}^3 B_{NS} } \right)^{1/8}= 6\, b^{1/8}
\tau^{3/8} \left( {\beta_{th} \over 0.3} \right)^{-3/8} \, {\rm eV}.
\label{epsilonc}
\end{equation} 
Thus, starting from far UV, quantum vacuum corrections to wave
polarization dominate over plasma contributions, so that normal modes
are linearly polarized and have similar resonant cross-sections
\citep{tlk02}.

\section{Radiation transfer}
\label{sec:main}

\subsection{General relations and Schwarzschild-Schuster method}

The photon transfer equation is best derived using covariant
formulation \citep{Lindquist}. 
The invariant
transfer equation for photon occupation numbers $n$ is 
\begin{equation}
k^\mu \partial_\mu n= \int S f_e  {d^3 p \over \gamma}
\end{equation}
where $S$ represents a collision rate between photons with occupation
numbers $n$ and electrons with a distribution function
$f$. Integration is over electron momenta $d^3p$, $\gamma$ is the
electron Lorentz factor, $k^\mu$ is the photon four-momentum.  We wish
to write down the equation of radiative transfer for resonant scattering
taking into account only linear terms in $\beta$. This is best done in
the frame where the electron is at rest and then transforming to the
fluid frame \citep{hsieh,bp82}.

The invariant collision integral is
\begin{eqnarray}
&&
S(\nu, {\bf l}; \epsilon , {\bf p}) = 
\nu_0 \int d\Omega'_0 \left( \left( { d \sigma ( \nu_0', {\bf l}'_0
 \rightarrow \nu_0, {\bf l}_0 ) \over d \Omega_0} \right) \left( {
 \nu_0' \over \nu_0 } \right)^2 \left( { d \nu_0' \over d \nu_0 }
 \right) [1+ n(\nu_0, {\bf l}_0)] n(\nu_0', {\bf l}'_0) \right. - \nn
 && \left.  \left( { d \sigma ( \nu_0 , {\bf l}_0 \rightarrow \nu_0',
 {\bf l}'_0) \over d \Omega_0} \right) n(\nu_0, {\bf l}_0) (1+
 n(\nu_0', {\bf l}'_0) \right)
\end{eqnarray}
where prime denotes a scattered photon, and subscript $0$ denotes
quantities measured in the electron's rest frame.

We are interested in terms linear in the thermal velocity of the
particles, and thus we neglect recoil effects.  In this case $\nu_0' =
\nu_0$ and the scattering cross-section is symmetric between the two
states:
\begin{equation} 
\left( { d \sigma ( \nu_0', {\bf l}'_0 \rightarrow \nu_0 {\bf l}_0 )
\over d \Omega_0} \right) = \left( { d \sigma ( \nu_0 , {\bf l}_0
\rightarrow \nu_0', {\bf l}'_0) \over d \Omega_0} \right) = { \pi^2
e^2 \over mc} \delta\left( \left\{
\begin{array}{l} \nu_0' \\ \nu_0 \end{array}
\right\}  - \nu_B \right)
(1+\mu_0^2)(1+\mu_0^{\prime,2})
\end{equation}
where $\mu_0$ is the cosine of the angle between the photon momentum and
the direction of the \Bf.  Neglecting induced process, $n \ll 1$, we
find
\begin{equation}
S= { \pi^2 e^2 \over mc} \nu_B \int f_e {d^3 p \over \gamma} \int
d\Omega'_0 (1+\mu_0^2)(1+\mu_0^{\prime,2}) \left( n(\nu_0', {\bf
l}'_0) \delta( \nu_0' - \nu_B) - n(\nu_0, {\bf l}_0) \delta( \nu_0 -
\nu_B) \right)
\end{equation}
Transforming to the lab frame, $\nu_0 = \gamma (1-\beta \mu) \nu$ and
using
\begin{equation}
n(\nu_0, {\bf l}'_0) = n(\nu', {\bf l}') \approx n(\nu,{\bf l}') +
\left( ({\bf l}'-{\bf l}) \cdot {\bf \beta} \right) \nu \partial_\nu
n(\nu,{\bf l}')
\end{equation}
 we arrive at
\begin{eqnarray}
 && k^\mu \partial_\mu n= \nu_B { \pi^2 e^2 \over mc} \int f_e {d^3 p
\over \gamma} (1+\mu_0^2 ) \int { d\Omega' } (1+\mu_0^{\prime,2})
\times \nn && \left( \left( 1+ \left( ({\bf l}'-{\bf l}) \cdot {\bf
\beta} \right) \nu \partial_\nu \right) n(\nu', \mu') \delta( \nu'
\gamma (1-\beta \mu')-\nu_B ) - n(\nu,\mu) \delta( \nu \gamma (1-\beta
\mu) - \nu_B ) \right)
\end{eqnarray}
where $\mu_0$ and $ \mu_0^{\prime}$ now should be expressed in terms
of laboratory frame quantities, $\mu_0 = (\mu-\beta)/(1-\beta \mu)$
and similarly for $\mu_0'$.  Assuming a non-relativistic one-dimensional
electron distribution, $f_e = f(p_{\parallel}) \delta(p_\perp)/ 2 \pi
p_\perp$, $\gamma \sim 1$, and a stationary plasma (so that odd power of
$\beta$ average to zero), we find 
\begin{eqnarray}
 && \mu \partial_x n = { \pi^2 e^2 \over mc} \nu_B \int f_e(\beta) d
\beta (1+\mu_0^2 ) \int { d\Omega' } (1+\mu_0^{\prime,2}) \times \nn
&& \left( n(\nu', \mu') \delta( \nu' \gamma (1-\beta \mu')-\nu_B ) -
n(\nu,\mu) \delta( \nu \gamma (1-\beta \mu) - \nu_B ) \right)
\label{f}
\end{eqnarray}

In order to investigate the salient features of resonant radiation
transfer, we make a simplifying one-dimensional approximation for
photon propagation (this is called the Schwarzschild-Schuster method
in radiation transfer). More advanced analytical methods, \eg, using
Gauss formula for numerical quadratures \citep{chand44} lead to
complicated equations. [Note that \cite{chan45} used Gauss method for
similar problem of line-driven wind, but neglected aberration effects
and kinematic enhancement of cross-section, both linear terms in
particle velocity $\beta$.]  In the Schwarzschild-Schuster method the
integral in the source function is replaced by a sum over forward
and backward propagating photons.  Eq.~(\ref{f}) is then replaced by a
system of two equations for $l =\pm 1$:
\begin{equation} 
l \partial_x n_l = { \pi^2 e^2 \over 2 m c} \nu_B \int f_e(\beta) d
 \beta \sum _{l' = \pm 1} \left( n(\nu', l' ) \delta( \nu' \gamma
 (1-\beta l' )-\nu_B ) - n(\nu,l) \delta( \nu \gamma (1-\beta l) -
 \nu_B ) \right)
	  \label{SD}
\end{equation}

\subsection{Radiative transfer in an inhomogeneous magnetic fields}

In an inhomogeneous \Bf radiation transfer occurs both in real space
as photons diffuse through the resonant layer, and in frequency space
as their frequencies are Doppler shifted at each scattering.
Neglecting the natural line width $\Gamma$,
at each scatter there is a conserved quantity, the velocity of the
particle
\begin{equation} 
\beta  = {\om - \om_B \over \om \cos \alpha} =
 {{\om}' - \om_B \over { \om}' \cos { \alpha}'}
\label{p}
\end{equation}
Thus, resonant radiation transfer on particles with different
velocities occur independently \citep{Zhel96}.  This allows a
simplification of the  problem if we consider $\beta$ instead of
$\om$ as an independent variable.  Accordingly, we transform to
variables $x-\beta$ by multiplying Eq.~(\ref{SD}) by
$\delta(\nu(1-\beta \mu) -\nu_B)$ and integrating over $\nu$:
\begin{eqnarray}
&& \left( \partial_x + { \partial_r \ln\nu_B } \partial_\beta \right)
n_+ = { \pi^2 e^2 \over 2 mc} f_e(\beta) \left( n_--n_+ + \beta (
n_-+n_+) \right) \nn && \left( - \partial_x + {\partial_r \ln\nu_B }
\partial_\beta \right) n_- = { \pi^2 e^2 \over 2 mc} f_e(\beta)
\left(n_--n_+ - \beta ( n_-+n_+) \right)
\label{system0}
\end{eqnarray}
where $n_\pm = \int d \nu n( \nu) \delta( \nu (1 \mp \beta) -\nu_B)$.

In a weakly inhomogeneous \Bf we can approximate $\om_B = \om_{B_0}
\left( 1-x/L_B\right)$.  After dimensionalizing, $x \rightarrow
x/L_B$, system (\ref{system0}) becomes
\begin{eqnarray} 
&& \left( \partial_x + \partial_\beta \right) n_+ = { \tau_{res}\over 2}
\left( n_--n_+ + \beta ( n_-+n_+) \right) \nn && \left( - \partial_x +
\partial_\beta \right) n_- = { \tau_{res}\over 2} \left(n_--n_+ - \beta (
n_-+n_+) \right)
 \label{system}
\end{eqnarray} 
where $\tau_{res}=L_B f_e(\beta) \pi^2 e^2/mc $.

System (\ref{system}) differs from the similar equations of
\cite{chan45} and \cite{Zhel96} due to the presence of velocity terms
on the right-hand side.  These terms come from the kinematic factor
$1\pm \beta$ in the collision rate.  Since we are interested in the
non-relativistic limit $\beta \leq 1$, in what follows we neglect
terms with $\beta$ on the right hand side. This allows us to obtain an
analytical solution for the transfer function. In appendix \ref{MC} we
verify the validity of this approximation using Monte-Carlo
simulations.  In the limit $\beta \rightarrow 0$ system (\ref{system})
becomes
\begin{eqnarray}
&& \left( \partial_x + \partial_\beta \right) n_+ = { \tau_{res}\over 2}
\left( n_--n_+\right) \nn && \left( - \partial_x + \partial_\beta
\right) n_- = { \tau_{res}\over 2} \left(n_--n_+ \right)
\label{Q2}
\end{eqnarray}

Resonant optical depth $\tau_{res}$ is generally a function of both $x$ (\eg
through the dependence of the density on $x$) and $\beta$. For
simplicity we assume that the density is constant, $\tau_{res}=
\tau(\beta)$.  Introducing new functions $f$ and $g$
\citep[\cf][]{chan45}
\begin{equation} 
n_+ = e^{-\int \tau(\beta) d \beta /2  } f
, \hskip .3 truein 
n_-  = e^{-\int \tau(\beta) d \beta /2 } g
\end{equation}
Eq.~(\ref{Q2}) becomes
\begin{equation}
\left(
\partial_x   +  \partial_\beta  \right) f = {\tau_{res}\over 2}  g,
\hskip .3 truein
\left(
\partial_x   -  \partial_\beta  \right) g = - {\tau_{res}\over 2}  f
\end{equation}
Which may be expressed as a second order partial differential equation
for $f$ only
\begin{equation}
\left(
\partial_x^2 -  \partial_\beta^2\right) f +
{ \partial_\beta \ln \tau_{res}} \left(
\partial_x   +  \partial_\beta  \right) f 
+{\tau^2 \over 4} f =0
\label{main}
\end{equation}
This equation describes the transfer of resonant cyclotron radiation
in a hot plasma in an inhomogeneous magnetic field.

As a simple analytically tractable case we assume that the
distribution of particles is a ``water-bag''
\begin{equation}
\tau(\beta) = \left\{ 
\begin{array}{cl} 
{\tau_0 \over 2 \beta_{th}}, & \mbox{ for $|\beta| < \beta_{th}$} \\ 
                       0, & \mbox{ otherwise.}
\end{array}\right.
\end{equation}
Eq.~(\ref{main})  then  becomes 
\footnote{This is  a wave equation with massive scalar field, 
or equivalently, a telegraphist equation.}
\footnote{\cite{Zhel96} have neglected the $\partial_x^2$ term inside
the resonance which is incorrect since the width of the resonance in
dimensionless $x$ is $\sim 2 \beta_{th}$, the same as in $\beta$}
\begin{equation}
\left(
\partial_x^2 -  \partial_\beta^2\right) f +
{\tau_0^2 \over 16 \beta_{th}^2 } f =0
\label{main2}
\end{equation}

Assuming  that $\omega_0 $ is the frequency resonant at $x=0$, $\om_0=\om_{B_0}$, the
resonance is given by 
$\om = \om_{B_0} ( 1- \xi) $ and $\om =  \om_{B_0} ( 1 +\eta)$.
where $\eta = \beta-x$, $\xi = x+\beta$ are forward and backward
propagating characteristics.
Changing to variables $\xi, \eta$, Eq.~(\ref{main2}) becomes
\begin{equation}
- 4 \partial_\xi \partial_\eta f+   {\tau_0^2 \over 16 \beta_{th}^2 }    f =0
\label{21}
\end{equation}
The Green's function for this equation is
\begin{equation} 
G(\xi-\xi_0, \eta-\eta_0)=
I_0\left( {\tau_0 \over  4 \beta_{th} } \sqrt{(\xi-\xi_0)(\eta-\eta_0)} \right)
\end{equation}
 
\begin{figure}
\plotone{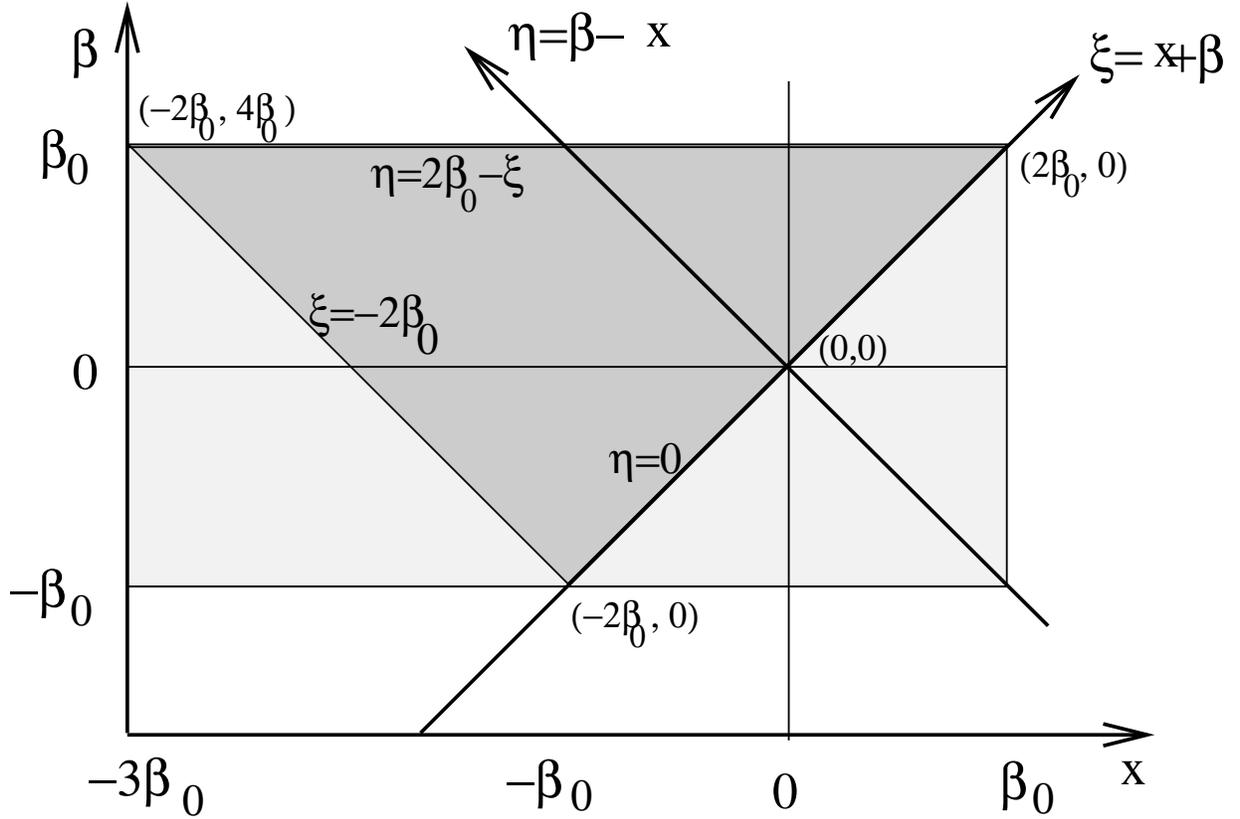}
\caption{Region of resonant interaction in $(x,\beta)$ and $(\xi,
\eta)$ coordinates 
($\eta = \beta-x$, $\xi = x+\beta$)  for a water-bag distribution.  Monochromatic light
with frequency $\om_0$  propagates through the resonance layer
centered at $\om_B(x=0)=\om_0$ and is subsequently scattered by
particles occupying the lightly shaded area $|\beta| <
\beta_{th}$. Resonance transfer occurs in the darkly shaded triangle
delimited by the lines $\eta=0$, $\xi=-2\beta_{th}$, $\eta=2\beta_{th}
-\xi$.  }
 \label{res-scat}
\end{figure}

The initial condition for this system is that a monochromatic wave
falls onto the inner edge of the resonant layer and that there is no
radiation falling on the outer edge.  In coordinates $\{\beta,x\}$ the
inner boundary of the resonant layer is located at {$\beta= -
\beta_{th}$, $x = - \beta_{th}$}, where {$\eta =0$, $\xi= -2 \beta_{th}$} (see
Fig.~\ref{res-scat}). At this point we have $n_+=
\delta(\om-\om_{B_0}) = \delta(\om_{B_0}(x-\beta)) = \delta(\eta)/
\om_{B_0}$.  And $n_-=0 \propto g \propto \partial_\xi f =0$.  Thus,
the two boundary conditions are
\begin{equation}
f(\xi= -2  \beta_{th}, \eta )  = \delta(\eta)
\hskip .3 truein
\partial_\xi  f(\xi= -2  \beta_{th}, \eta )=0
\label{22}
\end{equation}
where the intensity has been normalized to unity.

\begin{figure}
\plotone{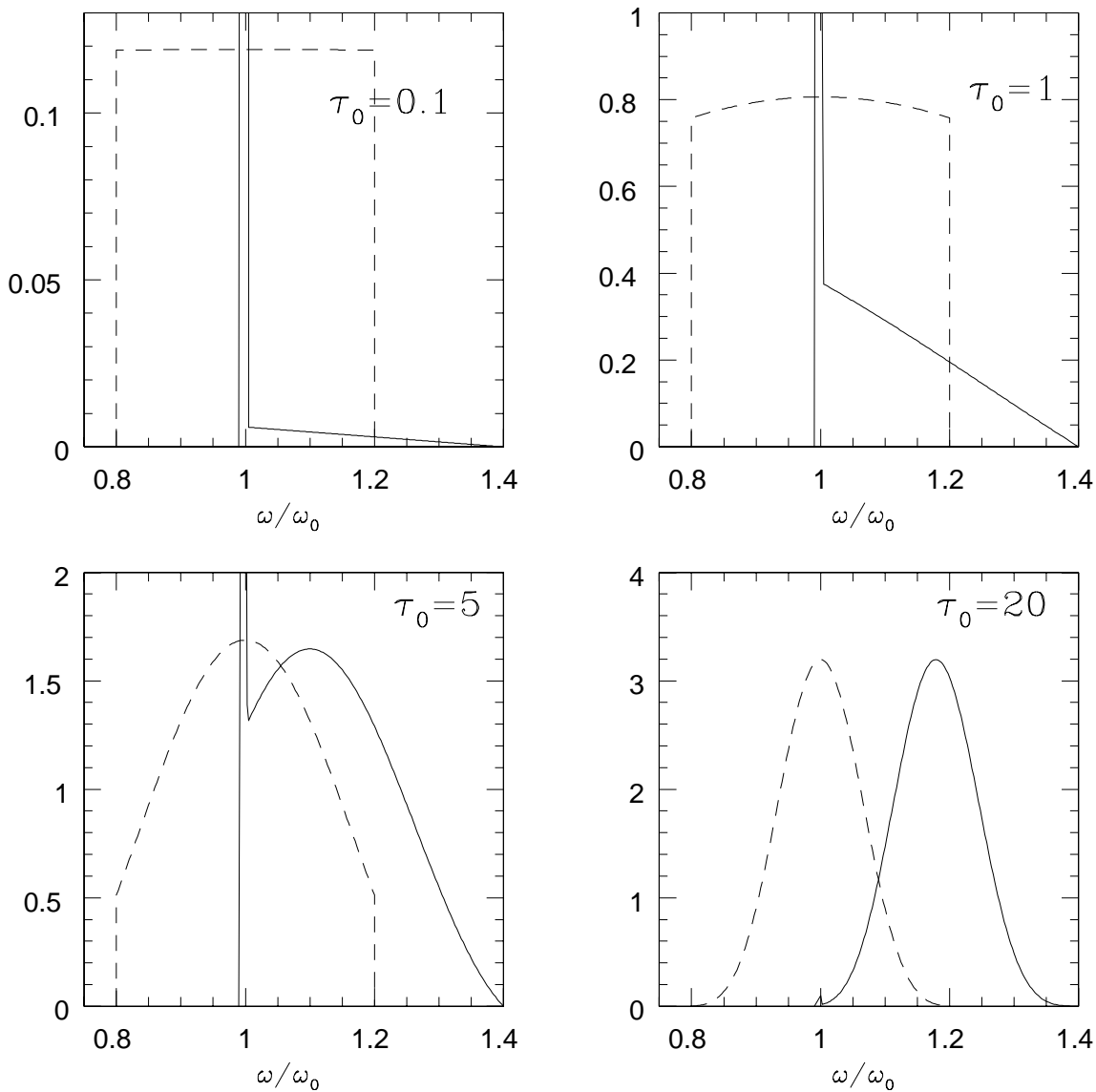}
\caption{Transmitted (solid lines) and reflected (dashed lines) fluxes
for a water-bag distribution of particles with $\beta_{th}=0.1$ and
different optical depths $\tau_0$.  For small $\tau_0 \leq 1$ the
reflected photons have a homogeneous distribution since the intensity
of the initial beam remains approximately constant. The reflected
intensity is $\propto \tau_0$.  The transmitted flux shows an initial
$\delta$-function and a double scattering tail $\propto \tau^2$, the
different panels show that as the optical depth increases the height
of the initial $\delta$-function peak decreases, while the intensity
of reflected beam increases.  In the limit $\tau_{res}\rightarrow \infty$
both the reflected and the transmitted intensities are equal (one half
of the initial intensity). For larger optical depth the dispersion of
both the reflected and transmitted intensities {\it decrease }.  In
this limit the transmitted photons are Doppler up-shifted on average
by $\Delta \om / \om = 1+2 \beta_{th}$, while the reflected photons have
on average $ \om \approx \om _0$.  }
\label{resscat}
\end{figure}

The solution of Eq.~(\ref{21}) given the boundary conditions
(\ref{22}) can be found using standard methods of mathematical physics
\citep[\eg][]{courant}.
\begin{equation}
f= \delta(\eta) + {\tau_0 \over 8 \beta_{th} } \sqrt{ \xi-\xi_0 \over \eta-\eta_0}
I_1 \left( {\tau_0 \over 4 \beta_{th} } \sqrt{(\xi-\xi_0)(\eta-\eta_0)} \right)
\end{equation}
and
\begin{eqnarray}
 && n_+ = e^{-(\eta+\xi+2\beta_{th})\tau_0/(8 \beta_{th}) } \left(
\delta(\eta) + {\tau_0 \over 8 \beta_{th} } \sqrt{ \xi-\xi_0 \over
\eta-\eta_0} I_1 \left( {\tau_0 \over 4 \beta_{th} }
\sqrt{(\xi-\xi_0)(\eta-\eta_0)} \right) \right) \nn && n_- = {\tau
\over 8 \beta_{th} } e^{-(\eta+\xi+2\beta_{th})\tau_0/(8 \beta_{th}) } I_0
\left( {\tau_0 \over 4 \beta_{th} } \sqrt{(\xi-\xi_0)(\eta-\eta_0)}
\right)
\label{01}
\end{eqnarray}
In the case of a water-bag distribution $\eta_0=0$, $\xi_0 = - 2 \beta_{th}$, 
and we find the transmitted and reflected fluxes 
\begin{eqnarray}
 && n_+ = e^{-
\tau_0/2} \left( \delta(\eta) + {\tau_0 \over 8 \beta_{th} } \sqrt{ 4
\beta_{th}-\eta \over \eta} I_1 \left( {\tau_0 \over 4 \beta_{th} }
\sqrt{\eta(4\beta_{th} -\eta)} \right) \right) \nn && n_-= {\tau_0 \over
4 \beta_{th}} e^{- \tau_0} I_0 \left( {\tau_0 \over 4 \beta_{th} }
\sqrt{(2\beta_{th}-\xi)(\xi+2\beta_{th}) } \right) 
\end{eqnarray}
where the transmitted
flux $n_+$ should be evaluated at $\xi=2\beta_{th} -\eta$, while 
the reflected flux $n_-$ at $\eta = 2\beta_{th} - \xi$.  Returning to
frequencies $\omega$, 
\begin{eqnarray}
 && n_+ = e^{- \tau_0/2} \left(
\delta(\om-\om_0) + {\tau_0 \over 8 \beta_{th} \om_0} \sqrt{ \om_0(1+4
\beta_{th})- \om \over \om-\om_0} I_1 \left( {\tau_0 \over 4 \beta_{th}
\om_0} \sqrt{(\om-\om_0)( \om_0(1+4 \beta_{th})- \om)} \right) \right)
\nn && n_-={\tau_0 \over 8 \beta_{th} \om_0} e^{- \tau_0/2} I_0 \left(
{\tau_0 \over 2 \beta_{th} \om_0} \sqrt{ \om_0(1+2\beta_{th})- \om)
(\om-\om_0(1-2 \beta_{th}))} \right)
\label{Main}
\end{eqnarray}
These relations are the solutions for the transfer function, giving
the fluxes of transmitted and reflected radiation for an initially
monochromatic wave (see Fig.~\ref{resscat}).


The reflection probability, found by integrating (\ref{Main}) over
frequencies, is
\begin{equation}
p_{refl} = \frac{1-e^{-\tau}} {2}
\label{pref}
\end{equation}
In the large optical depth limit $p_{refl}=1/2$.  \citep{Zhel96}.

For small optical depths 
\begin{eqnarray}
 && n_+ = \delta(\eta) \left( 1-{ \tau_0
\over 4 \beta_{th}} (\xi+2\beta_{th}) \right) \nn && n_- = { \tau_0 \over 4
\beta_{th} } 
\end{eqnarray}
 while in the limit of large optical depths
\begin{eqnarray} 
&&
n_+ = { \sqrt{\tau_0 \over \beta_{th} } \over 4 \sqrt{2 \pi}  \om_0 } 
\left(1+4 \beta_{th} - {\om \over \om_0} \right)^{1/4}
\left( {\om \over \om_0} -1 \right) ^{-3/4} 
e^{  (\om - \om_0 (1+ 2 \beta_{th}) )^2 \tau_0 / ( 16 \beta_{th}^2 \om_0^2)  }
\nn && 
n_-= { \sqrt{\tau_0 \over \beta_{th} } \over 4 \sqrt{2 \pi}  \om_0 }
\left( (1+2 \beta_{th} -  {\om \over \om_0})({\om \over \om_0}-(1-2 \beta_{th}))) \right)^{-1/4}
e^{ (\om - \om_0)^2 \tau_0 / ( 16 \beta_{th}^2 \om_0^2)  } 
\label{ans}
\end{eqnarray}

\subsection{Qualitative properties of resonant transfer function}

The properties of the transmitted and reflected fluxes (\ref{Main})
are very different from Thomson scattering and from resonant cyclotron
scattering in a homogeneous \Bf. (Fig.~\ref{resscat}).  In the small
optical depth limit, $\tau_0 \ll 1$, the reflected wave has an
approximately constant spectral flux, $\propto \tau_0$, centered at
$\om_0$ with dispersion $\Delta \om/\om \sim 2 \beta_{th}$. The
transmitted radiation, $\propto \tau_0^2$, is up-scattered, spanning
$\om_0 < \om < (1+ 4\beta_{th}) \om_0$.

In the large optical depth limit $\tau_0 \gg 1$ the spectral fluxes of
the reflected and the transmitted radiation are equal, so that a
resonant layer with high optical depth is half opaque,
Eq.~(\ref{pref}), see also \cite{Zhel96}.  This is in sharp contrast
to the case of non-resonant scattering and to the case of resonant
cyclotron scattering in a homogeneous \Bf, when the transmitted wave
is only $\propto e^{-\tau} \ll 1$.  Half opaque resonant layers are
also reminiscent of the case of line-driven winds \citep{sobo47}.
{\it The transmitted radiation is Compton up-scattered by a factor
$\sim 1+ 2 \beta_{th}$, while the reflected waves on average have $\om
\sim \om_0$}. Note that this up-scattering is done by a
non-relativistic, static warm medium in the limit where recoil effects
are neglected, \cite[\cf][]{Komp}.  Next, {\it as optical depth
increases, the transmitted and the reflected fluxes have increasingly
narrower distributions. } The typical dispersion around the central
frequency is $\Delta \om/\om_0 \sim 2 \sqrt{2} \beta_{th} / \sqrt{\tau}$
for $\tau_{res}\gg 1$.  This again can be contrasted with non-resonant
Thomson scattering, in which case dispersion increases with increasing
$\tau_{res}$.

\begin{figure}
\plotone{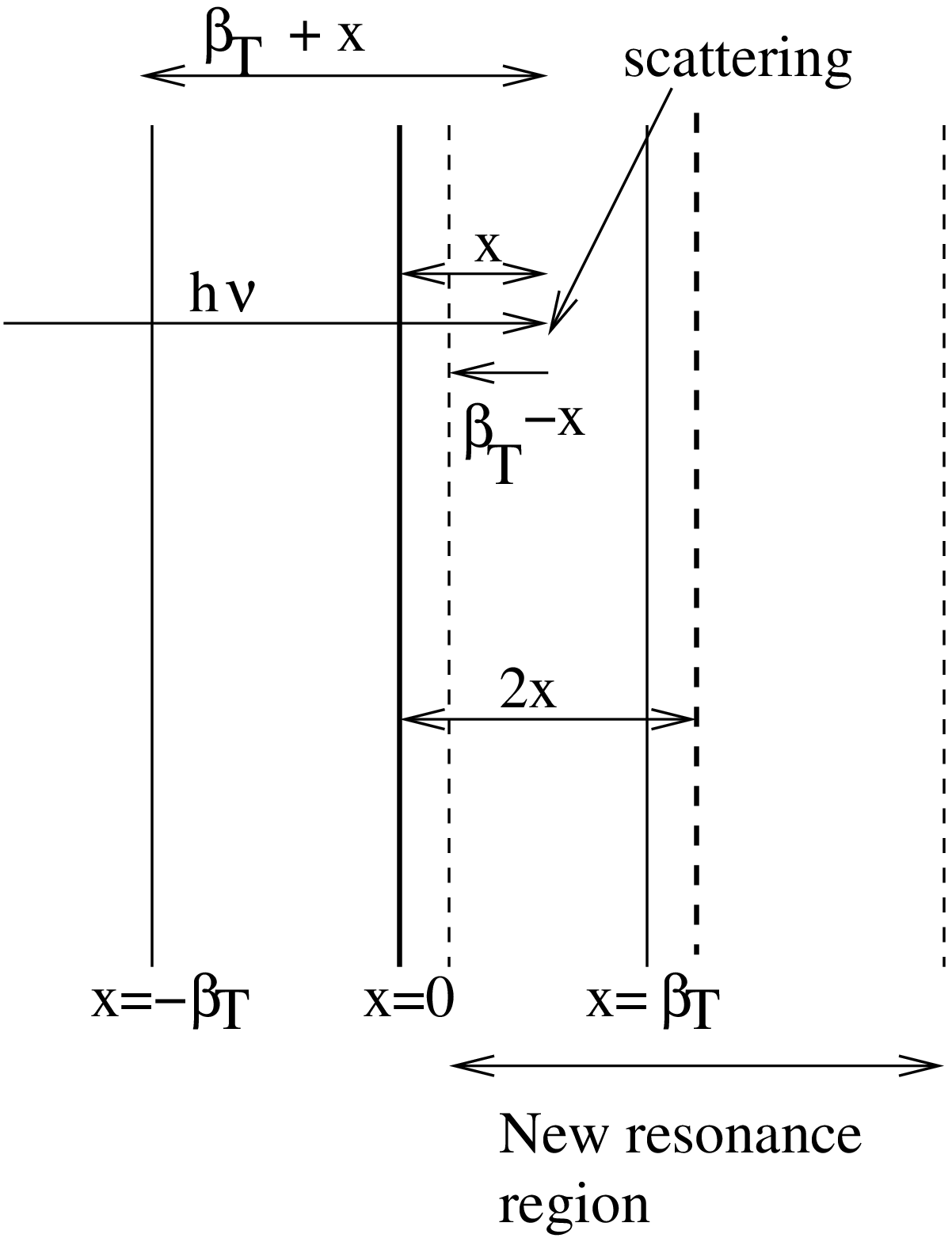}
\caption{Illustration why photon always
spends  dimensionless time $2 \beta_{th}  $ in the resonance region
($2 \beta_{th}  L_B$ in regular units, where $L_B$ is the \Bf inhomogenuity scale) and  why resonant layer is semi-opaque in the limit of high optical
depth. Initially, resonant surface $\om_0=\om_B$ is located at $x=0$. A photon experiences backward scattering at position $x$ after traveling a distance $\beta_{th} + x $ inside
a resonance layer. New resonance layer is centered at $2 x  $
so that a photon  is located a distance
 $\beta_{th} -x$ from the inner edge of the resonant layer. If it escapes   without furher scattering, it would have spent total time $2\beta_{th} $ inside a resonant layer.
For $\tau_0 \gg 1$ an incident photon  is
scattered at $x \sim - \beta_{th}$ on electrons with $\beta \sim -
\beta_{th}$.  Half of the photons are then re-emitted forward with
$\om=\om_0$, and half are re-emitted backward with $ \om=\om_0(1+2
\beta_{th})$. For backward propagating photons the new resonant surface is
located at $x \sim - 2 \beta_{th}$, so that photons are trapped between 
resonant layers and  escape with approximately
equal probability  in both directions.
 }
 \label{2betaT}
 \end{figure}

Consideration of  photon  residence time  inside the resonance layer
and number of scattering in the limit of high optical depth also
bears a lot of surprises.
 {\it Residence time inside the  layer
is  independent of optical depth}, while number of scattering 
scales as $\sim \tau_{res}$!
To understand this, consider a photon moving outward through a resonant layer.
 Suppose it is scattered backward
  at a point $-\beta_T<x<\beta_T$ on an electron with resonant
  velocity $\beta= x$ (in dimensionless units). 
  It has already traveled a distance $\beta_T + x$ inside
  a resonance layer. After the 
  scattering the photon energy is $\propto 1+ 2 x$  and  the new resonant
  surface is located at $2 x$, so that during scattering the photon is located a distance
   $\beta_T -x$ from the edge of the 
   resonant layer. If it leaves without any further scattering
   it has traveled a total distance
   $\beta_T + x + \beta_T -x = 2 \beta_T$ inside a resonant layer
   (see Fig. \ref{2betaT}).
   It is easy to see that this quantity 
   does not change at any scattering is valid for any number of
   scatterings.
Thus, 
regardless of $\tau_{res}$  a photon spends a time
$2 \beta_{th}  r_{res} / c$  in the region where 
it can resonate with electrons.
This can be compared with scaling $\propto \tau$ for  residence time and
 $\propto \tau^2$ for  number of scattering for non-resonant scattering.
An initial short $\delta(t)$-pulse will, of course, be smeared and delayed
since depending on scattering
history a photon leaves the resonant layer at
different spacial coordinates. In the limit of large optical depth
spectral smearing is negligible, so that
{\it  all transmitted photons are delayed by $\beta_{th}  r_{res} / c$ }
with respect to freely propagating ones.
(reflected photons are, equivalently, being reflected by a ''wall''
located at  $ r_{res} (1- \beta_{th})$).

The two most unusual characteristics of resonant cyclotron scattering
in a strongly inhomogeneous \Bf are the facts that in the high optical
depth limit the transmission coefficient is 1/2 and that the
transmitted multiple-scattered photons are preferentially up-shifted
in energy (for a water-bag distribution they are {\it always}
up-shifted in energy).  According to \citep{Zhel96},
semitransparence of the resonant layer
 can be explained in the following way.
 In the high optical depth limit, most photons will
experience the first scatter on electrons moving backwards (along the
negative direction $x$) with $\beta \sim -\beta_{th}$. This occurs near
the dimensionless coordinate $x \sim -\beta_{th}$.  Photons that are
re-emitted backwards have their frequency increased by $1+2\beta_{th}$.
For these photons the resonant layer is located closer to the star,
shifted by $\Delta x = - 2\beta_{th}$, so that the reflected photon finds
itself outside of the resonant layer. Thus, the photons are trapped
between two resonant layers.  As a result their escape probability is
1/2 in each direction (see Fig.~\ref{2betaT}).

To understand the energy up-shift of the transmitted
multiple-scattered photons, suppose that a photon is scattered
backward at some position $x$ on a particle with velocity
$\beta_1$. The energy of the photon increases for $\beta_1<0$ or
decreases for $\beta_1>0$.  The photon can resonate again only with
particles whose velocity satisfies $\beta > \beta_1$. As a result, in
the subsequent forward scatter its energy will either decrease by an
amount smaller than the initial increase (for $\beta_1<0$), or
increase by an amount larger than the initial decrease (for
$\beta_1>0$).

\subsection{Validity of approximations}

For non-relativistic plasma, $\beta_{th} \ll 1$, and for low energy photons, 
$\epsilon  \ll m c^2$, recoil frequency shift  is small in  each
scattering event. Since for resonant scattering number of scattering in
the limit $\tau_{res}\gg 1$   scales as $\sim \tau_{res}$, the  condition for
neglect of accumulated
recoil is ${ \epsilon \over m c^2} \tau_{res}\ll \beta_{th} $ where
$ \epsilon  $ is  photon energy. 
For $\sim$ keV photons and $\beta_{th}  \sim 0.1$,
this limits optical depth to $\tau_{res} \leq 10$.
 Since
 in this case the  number of scatterings increases with optical depth slower
 than  for non-resonant scattering,  this condition on optical depth is less
 stringent.

For non-relativistic plasma temperatures transition to 
 higher harmonics $s$
 are suppressed by $ \beta_{th}^{2s} \ll 1$.
 At high optical depths, $\tau_0 \gg 1$, such transition may still contribute
 efficiently to radiation transfer in \Bf 
 (similar to the case of energy tranfer
 in wings of highly absorbed lines). The conditions to neglect
 higher order transition is then $ \tau_0 \beta_{th}^{2s} \ll 1$.

In case of non-resonant scattering,
 for high optical depth number density of photons increases if compared
 with the case of no absorption. This leads to increased importance of 
 induced processes. For resonant scattering 
 this is relatively  small effect since increase in photon residence
 time (and corresponding number density)
 inside the magnetosphere is small, typically a factor
 $1+\beta_{th} \approx 1$. 

\section{Magnetar spectra due to resonant scattering}

\subsection{Magnetospheric plasma in magnetars}

In magnetars, large scale currents flow in the magnetosphere, pushed
out of the neutron star interior by the slow unwinding of the non-potential
magnetic field \citep{tlk02}. The strength of the electrical currents
is determined by the speed with which the internal magnetic field
unwinds.  It is expected that the resulting toroidal magnetic field
$B_\phi$ is limited by the poloidal field $B_p$, $B_\phi\leq B_p$ (for
larger fields the configuration is likely to be unstable to, \eg, kink
modes). If $ B_\phi \sim \Delta \phi B_p$, where $\Delta \phi \leq 1$
is a pitch angle of the magnetic field, the particle density required
to carry the current will result in a plasma frequency of the order
\begin{equation} 
\om_p^2 \sim \om_B {c \over v} {c \over r} \Delta \phi
\end{equation}
where $v \leq c $ is a the typical velocity the current carries.  The
resulting optical depth
\begin{equation}
\tau_{res}\sim {c \over v} \Delta \phi
\end{equation}
is of the order of unity, regardless of the mass, frequency and
magnetic field \citep{tlk02}.  Thus, currents flowing in magnetar
magnetospheres may provide large optical depth to resonant cyclotron
scattering.  Electrons then can be heated due to the development of a
two stream instability operating on short time scales, $\sim 1/\om_p
\sim \sqrt{r / \om_B c \tau}$.

\subsection{Multiple scattering of radiation in the  X-ray band}

Emission from the neutron star's surface may be modified by resonant
scattering.  As radiation from the surface propagates in the
magnetosphere it is scattered in a resonant layer, some of it is
reflected back toward the star and may experience multiple scattering
between different hemispheres, so that the emergent spectrum is a sum
over many reflections. At every reflection from a resonant layer
radiation is broadened by the random motion of resonant electrons,
without changing the mean frequency, while the multiply scattered
transmitted radiation is Compton up-scattered $1+2 \beta_{th}$.  If the
scattering plasma occupies a large fraction of the magnetosphere, the
resulting spectrum is a sum of broadened Plank spectra (the broader
the component, the smaller its relative intensity), Compton
up-scattered by an optical depth-dependent factor of the order $1+2
\beta_{th}$.

First, we calculate semi-analytically the  transmitted flux for initially
Plankian spectra by summing over multiple reflections between resonant
layers located in different hemispheres. In the spirit of our one
dimensional approach, we assume that the scattering surfaces are plane
parallel.  If the spectrum of the surface emission is given by
$n_s(\om_0)$ then the observed spectrum becomes
\begin{equation}
n_{obs} = \int d \om_0 n_+ (\om, \om_0) n_s(\om_0)  +
 \int d \om_1 n_+ (\om, \om_1) \int n_- (\om_1, \om_0) n_s(\om_0)  d \om_0 
+ \cdots
\end{equation}
where the sum is over multiple reflections. Fortunately, the sum
quickly converges since each consecutive term is at least two times
smaller than the previous one (for numerical computation we limit the
sum to 6 terms, which results in an error of the order of $2\%$ at
most).

\begin{figure}
\plotone{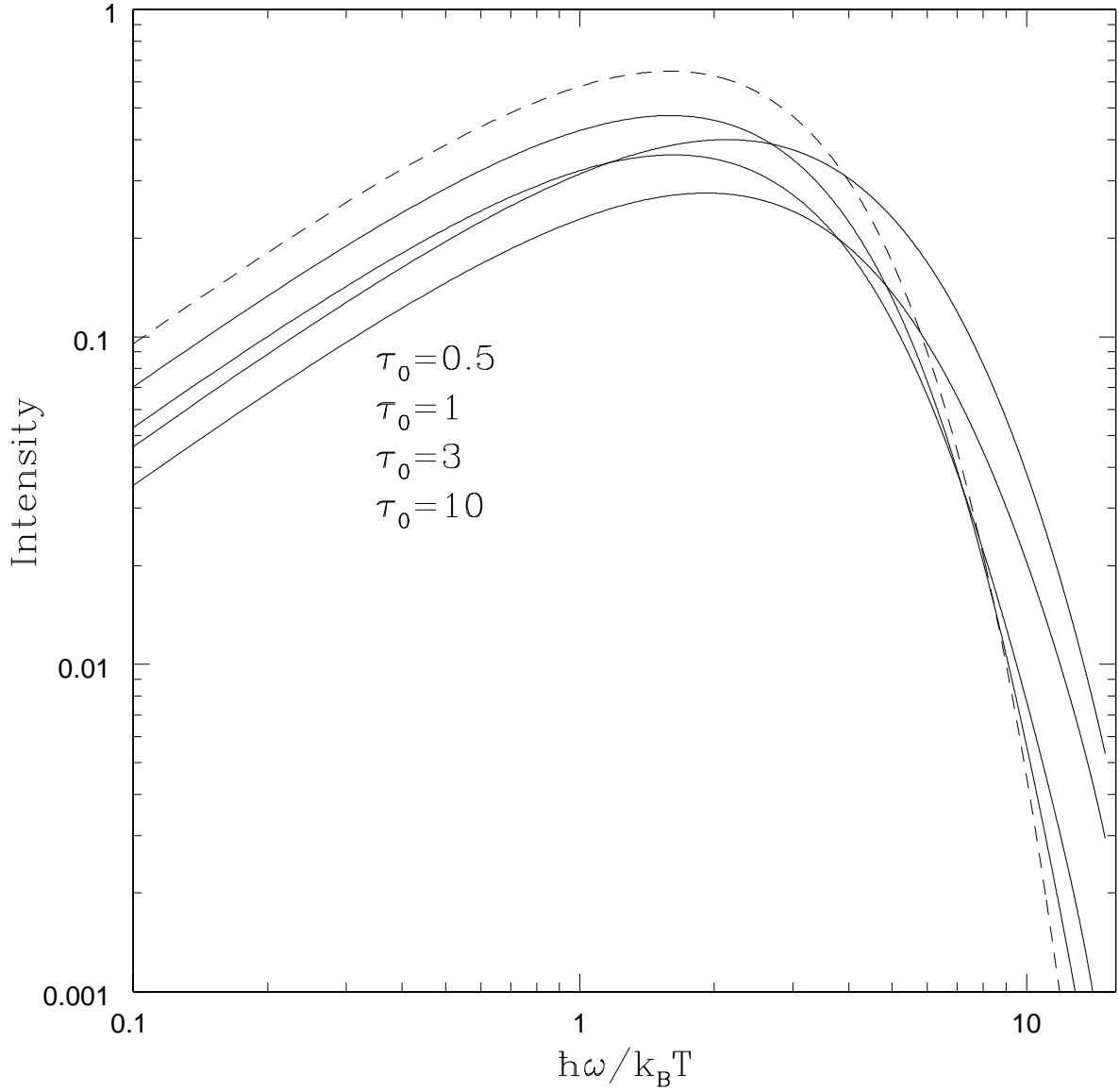}
\caption{Modification of an initially Plankian spectrum (dashed line)
by multiple cyclotron scattering; $\beta_{th} =0.3$ for different values
of $\tau_{res}$.}
\label{iout}
\end{figure}

Results of the calculations are presented in Fig.~\ref{iout}.  First,
the spectrum is overall up-scattered.  
  {\it The initial spectrum is shifted to higher
energies}, by as much as $1+2\beta_{th}$ in the limit $\tau_0 \gg 1$ (for
smaller optical depths the shift of the emission peak is smaller).  As
a consequence, the real surface temperature is smaller by as much as a
factor $1+2\beta_{th}$ than the one inferred from the Compton
up-scattered flux.

\begin{figure}
\plotone{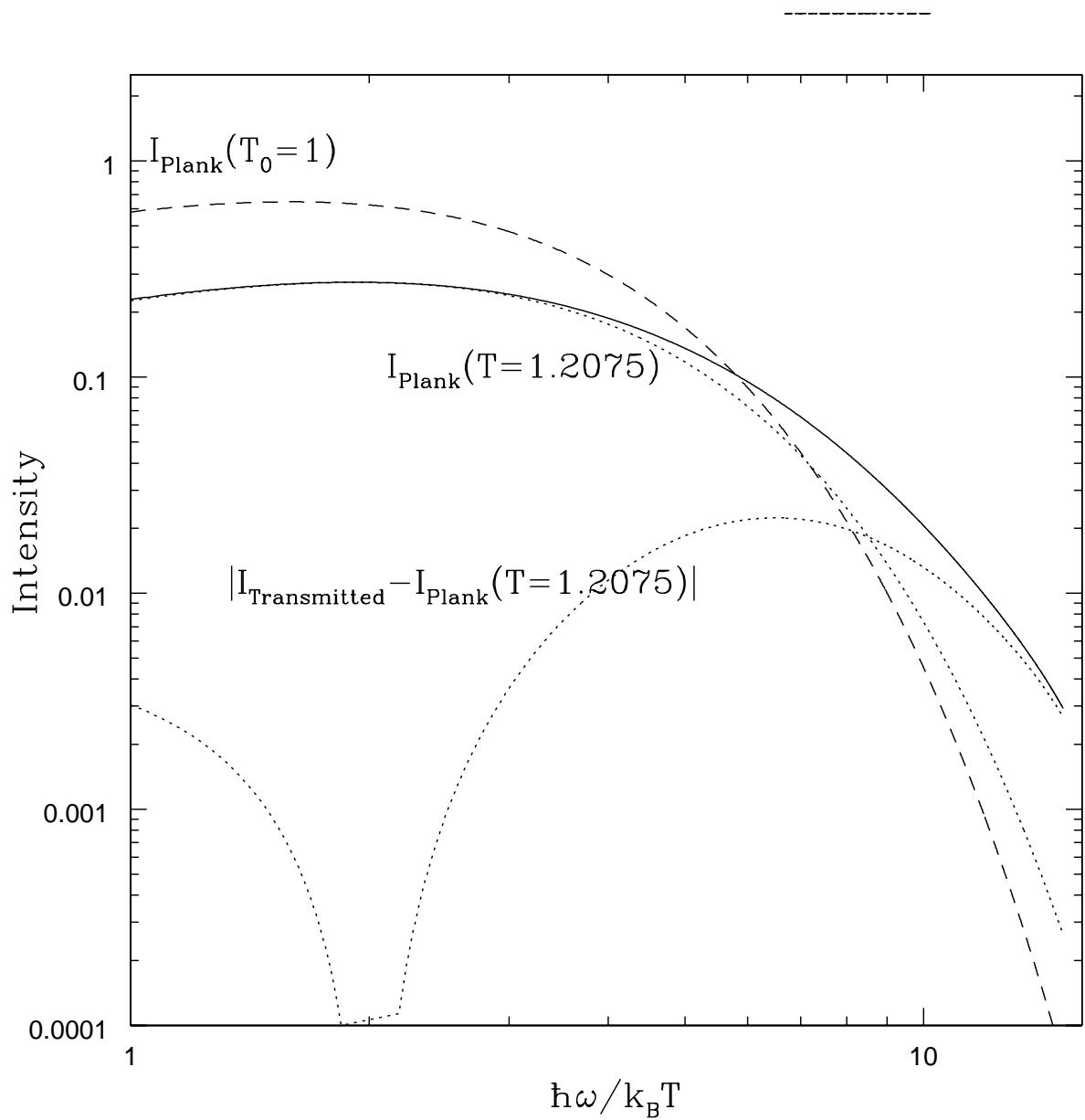}
\caption{A fit to the transmitted spectrum (solid line) for optical
depth $\tau_0=3$ and $\beta_{th} =0.3$.  Dashed line is the initial Plank
spectrum with $T_0=1$, Dotted lines are a Plank fit to the transmitted
spectrum at the peak of the emission, $T=1.25$ and a residue. The residue
is strongly non-thermal. 
}
\label{itau}
\end{figure}

Secondly, {\it for optical depth $\tau_0 \sim 1$ the transmitted
spectrum is non-Plankian}, see Fig.~\ref{itau}.  Modification of the
spectra comes both from increased photon dispersion during each
reflection as well as from summing contributions from several
reflections.  Note, that the non-Plankian form of the transmitted
spectrum is mostly prominent at $\tau_0 \sim$ few.  In this case the
reflected flux becomes comparable to the initial flux.  On the other
hand, in the limit of large optical depth $\tau_0 \gg 1$, each
reflection {\it does not lead to an increase in the flux dispersion},
so that the transmitted flux remains mostly Plankian, but with the
peak shifted to larger energies by $1+2\beta_{th}$.

The asymptotic spectrum at frequencies much higher than the
initial frequency can be estimated if we  assume that each reflection acts as a
Gaussian filter with a typical dispersion $\sigma \sim \beta_{th} \om_0
\leq \om_0$ and that the reflection probability is $0< p< 1/2$.
Neglecting the small frequency shift during the final transmission,
the resulting flux $F$ is a sum over all reflections
\begin{equation} 
F \propto \sum_{n=0}^{\infty} {e^{- {\om^2 \over 2 (n+1) \sigma^2}}
 p^{n+1} \over \sqrt{ 2 \pi (n+1) } \sigma }.
\end{equation}
For $n\gg 1$ we can substitute the sum by an integration and evaluate
the integral by the steepest decent method at $n \sim \om/( \sqrt{ 2
\ln (1/p)} \sigma)$,
\begin{equation}
F \propto {1 \over \sqrt{2} \sigma} e^{- \sqrt{2 \over \ln (1/p) } {
 \om \over \sigma} }
\end{equation}
Qualitatively, it takes on average $n \sim \om/(\beta_{th} \om_0)$
scatterings to reach a frequency $\om \gg \om_0$, but the intensity of
radiation is decreased by $p^{n+1}$.  Thus, scattering by weakly
relativistic warm plasma with $\beta_{th} \leq 1$ produces a non-thermal
spectrum with an exponential cut-off at $\om\sim k_BT/(\hbar \beta_{th})$.

\subsection{Spectral Fitting}
\label{sec:spectral fits}

Motivated by the fact that resonant Compton scattering qualitatively
reproduces magnetar spectra, we performed a preliminary analysis
to verify that it can also do so quantitatively.  We decided to fit
our model to a 28.8~ks long \cxo\ X-ray observatory observation of the
anomalous X-ray pulsar (AXP) \tfe.
The data were taken with the ACIS detectors (S3 chip) in \texttt{CC mode}.
In the literature \tfe's spectrum
has been characterized by a two-component model consisting of a
blackbody of temperature $k_BT \sim 0.6$~keV and a power-law tail with
spectral index $\Gamma\sim 3$. We chose this source because its
spectrum is intermittent between the anomalous X-ray pulsars (AXPs)
and the soft-gamma repeaters (SGRs). An analysis of all magnetar
spectra will be  done in the future.

Given that our model is semi-analytical we created a table of
intensities for different values of $\beta_{th}$, $\tau_{res}$ and $k_BT$.  We
developed software to convert this grid into a format that could be
read by the X-ray spectral fitting program
\texttt{XSPEC}\footnote{http://xspec.gsfc.nasa.gov}. Using
\texttt{XSPEC} the model was folded through the instrumental response
and multiplied by a photoelectric absorption model. Only counts between
0.5-1.67 and 1.91-6.7~keV were included in the fits, leaving 281
channels for spectral fitting. The 1.67-1.91~keV region of the
spectrum was ignored because of an instrumental feature in the
spectrum. For more details on the analysis of this \cxo\ observation
of  \tfe\  see \citep{gkw+05}.  In our fits
$\beta_{th}$ and $\tau_{res}$ were held fixed at predetermined values but we
allowed the column density $N_H$, the temperature $k_BT$ and the
normalization to vary freely. We repeated the procedure for different
values of $\beta_{th}$ and $\tau_{res}$ until we minimized a $\chi^2$
statistic. Even with our course grid in $\beta$ and $\tau_{res}$ we were
able to successfully reproduce the observed spectrum.  The
fit had a reduced $\chi^2_{\mathrm{dof}} =1.05$ for $278$ degrees of freedom (dof).
We compared our model to a blackbody plus power law model and a
double blackbody model. A summary of our spectral fits is given in
Table~\ref{ta:spectra}.  

\begin{figure}
\plottwo{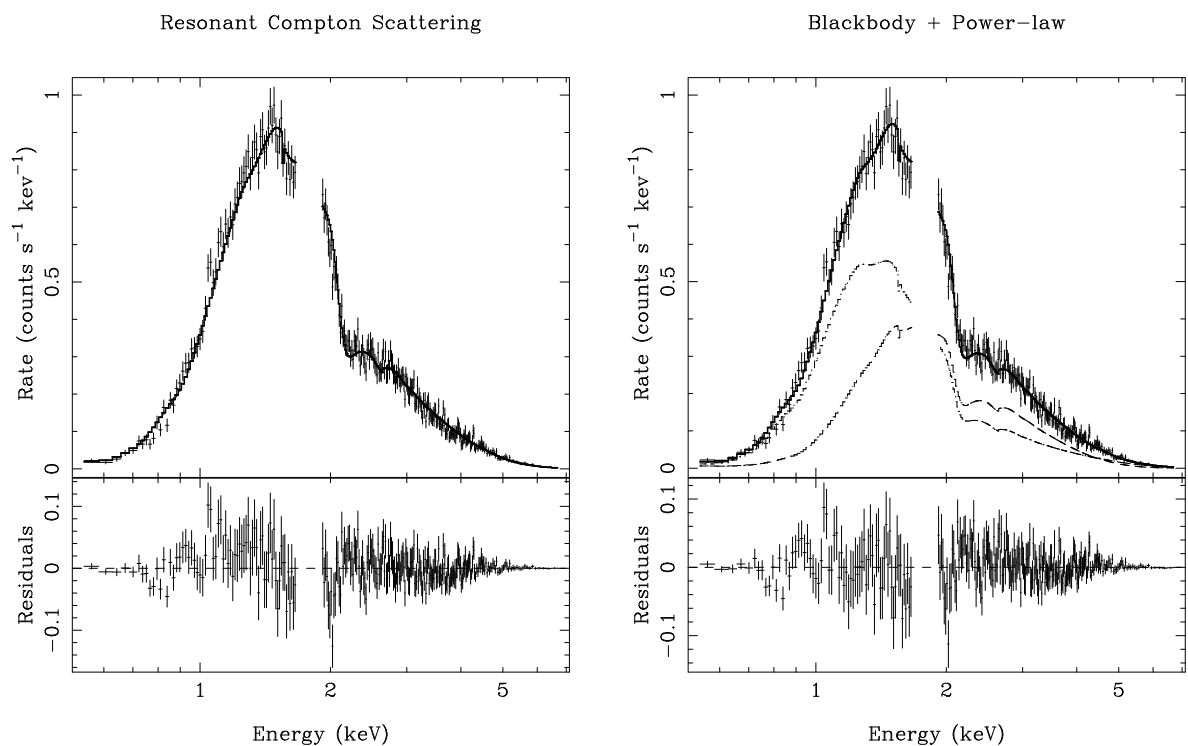}{spectra_bbody_po.eps}
\label{fig:fits}
\caption{X-ray Spectrum of the Anomalous X-ray Pulsar \tfe\ as
observed by the \cxo\ observatory. Left: The data fit with our
resonant Compton scattering model. See Table~\ref{ta:spectra} for
best-fit parameters. Right: Same spectra but fit to a blackbody plus
power law. see Table~\ref{ta:spectra} for the best-fit parameters. The
dashed and dotted lines represent the individual model components. See
text for details on the analysis}
\end{figure}

For comparison we plot the best-fit blackbody plus power law model
and the best-fit resonant Compton scattering model for this
observation in Figure~\ref{fig:fits}.  It is interesting to note that
our model fairs just as well as the ``canonical'' magnetar spectral
model of a blackbody plus power-law model (see
Table~\ref{ta:spectra}). Furthermore, it is interesting to note that
all three models listed in Table~\ref{ta:spectra} have the same number
of parameters and fit the data well, however, the resonant Compton
scattering model has an advantage: it is a single (rather than a
double) component model which has a clear physical interpretation
within the framework of the magnetar model.



\subsection{Secular spectral variations in magnetars}

Most magnetar candidates (AXPs and SGRs) show evidence of secular
spectral variability \cite[\eg][]{marsden01,wt04}. In the magnetar
model this is readily explained by a variable current flowing in the
magnetosphere producing variable optical depth to resonant scattering
\citep{tlk02}.
\footnote{ Isolated neutron star RX J0720.4-3125 also shows spectral variations
of its {\it black-body component } on time scale of years
\cite{Paerels01,devries04}. Resonant scattering may offer a possible
explanation, but an upper limit on the \Bf $\leq 3 \times 10^{13}$ G
\cite{kaplan02} seem to exclude the magnetar hypothesis for it.}
Spectral
variations in the persistent emission of  magnetar candidates 
have been prominent during
active bursting periods. For example, significant spectral evolution
was observed during an outburst of AXP \tfn\ involving $\sim$80 bursts
within a 3~hr observations \citep{kgw+03}.
More recently a similar, yet much more subtle change was reported by
\citet{roz+05} in AXP \soe. They find that the source's spectrum has
been softening, specifically they find that the photon index increased
from $\Gamma\sim 2.4$ to $\sim$2.8. 
In both cases the  correlation between the
luminosity and the hardness of the non-thermal component of the
source's spectrum is consistent with the large-scale current bearing
magnetosphere model of \citet{tlk02}. 

\section{Other  applications}

\subsection{Smearing of spectral lines}

\begin{figure}
\plotone{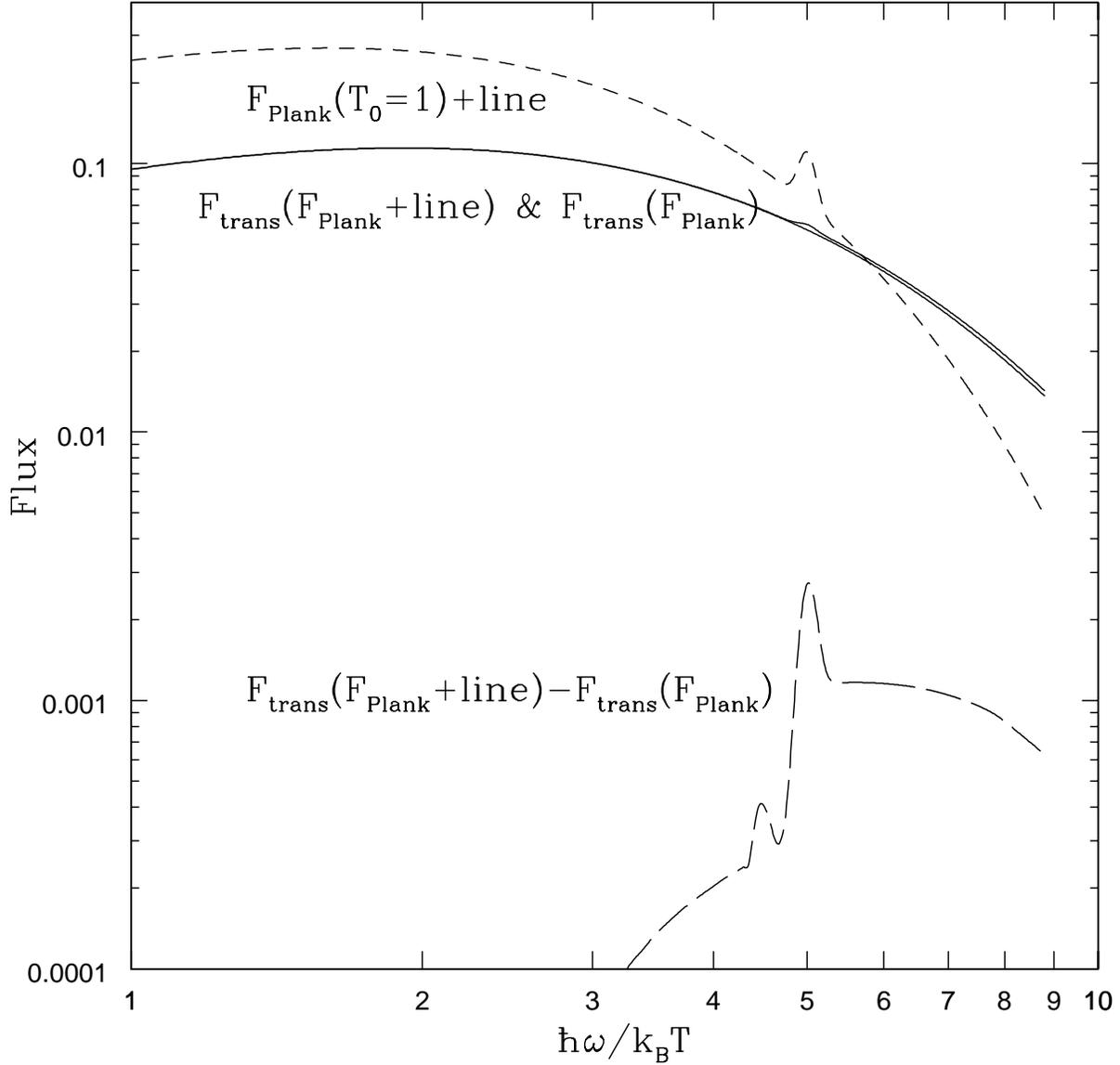}
\caption{Effects of resonant scattering on a line in a spectrum.
Initial spectrum (dashed line) is Plankian plus an emission line
located at $\hbar \omega_{line} = 5 k_B T$, with a width $\Delta
\omega_{line} =0.1 k_B T$ and total intensity 1/100 of the black
body. The transmitted intensity (solid line) is barely above the
transmitted intensity for the no line case, Fig.~\ref{iout}, (dotted
line). The difference between the two (long dashed line) has a
complicated profile, but it's intensity is only $\sim 1\%$. }
\label{iline}
\end{figure}

Atmospheres of neutrons stars were expected to produce spectra
abundant with spectral features \cite[\eg][]{zavp02}.  Though spectral
features were observed in some pulsars, their absence in some
prominent sources like \efs\ \citep{drake02}, poses a challenge to
atmospheric models, \cite[see, \eg,][for review]{kerk03}.  Resonant
scattering may provide a possible explanation.  Let us calculate the
effects of cyclotron scattering in a magnetosphere filled with plasma
on a narrow spectral feature superimposed on a thermal spectrum.  The
initial spectrum is taken to be a thermal plus a narrow emission line
located at the energy $\epsilon_{line} = 5 k_B T$ with a width $\Delta
\epsilon_{line} =0.1 k_B T$ and total intensity 1/100 of the 
blackbody (Fig.~\ref{iline}). The emission passes through the magnetosphere
with optical depth $\tau_0=1$. After multiple reflections the
transmitted intensity is virtually indistinguishable from the one
obtained from just a thermal spectrum without an emission line.  Thus,
scattering in the magnetosphere may provide an explanation for the
lack of spectral features in neutron star.

\subsection{Optical  emission}

The character of resonant scattering 
changes  for UV and longer wavelengths, when the wave
polarization at resonance becomes circular
Eq.~(\ref{epsilonc}). Consider, for simplicity, the scattering of
unpolarized light in a monopole \Bf in a one-dimensional approximation
 in the case when normal modes are circularly polarized.
 For unpolarized light propagating through an
electron-ion plasma, half of the intensity (with circular polarization
such that electric field of the waves rotates in the same direction as
the positive charges) is transfered without scattering, while the
other half (with polarization such that the electric field rotates
with the electrons) is scattered with the transfer functions
calculated in Section \ref{sec:main}.  In the limit $\tau_{res}\gg 1$ half
of the scattered radiation is re-emitted forward, half backward, so
that the transmission coefficient is $3/4$ instead of $1/2$ for the
case of linearly polarized waves. 

In addition, 
two possibilities exist for multiple scattering of a reflected photon
from the  opposite hemisphere. If a photon
propagates mostly outside the polarization limiting radius,
its polarization does not change along the
  ray, while for propagation inside the polarization limiting radius
  its polarization  evolves adiabatically. 
These two possibilities  result in qualitatively different scattering regimes:
if a photon keeps its polarization, then in the  simple one-dimensional
picture
in the opposite hemisphere the electric field of the wave
rotates in the opposite direction to that of the local electron
so that photon will not experience any more
scatterings.
On the other hand, if  polarization evolves  adiabatically, by the time
a photon would reach a resonance in the opposite hemisphere it will become
virtually unpolarized and thus can be scattered. 

Location of polarization  limiting  radius $ r_{lp}$  can be estimated from
the condition
$\Delta n r_{lp} \sim \lambda$ where $\lambda$ is wavelength and
$\Delta n$ is the difference between refractive indexes of the two normal modes. In the relevant regime $\Delta n = {2\alpha \over 60 \pi} (B/B_{QED})^2$
\citep{heyl97}, which gives
\be
 r_{lp} = \left({b R_{NS} \alpha_{fin} \epsilon
  \over 60 \pi^2  c \hbar} \right)^{1/5}
   =  5.7 \times 10^7 b^{2/5} \, {\rm cm}
\ee
where $\alpha_{fin}$ is  fine structure constant.
This is typically larger than resonance radius, 
so that  polarization varies adiabatically.

\subsection{Young pulsars}

The importance of cyclotron scattering in the magnetosphere on the
soft X-ray properties of young pulsars has been first pointed out by
\cite{wrhz98}, but see also \cite{rude03}.  It was suggested that abundant
pair creation may occur on closed field lines which would produce a
large optical depth to cyclotron scattering. Unfortunately, the model
of the cyclotron blanket proposed in the above references is incorrect
since it was assumed that in the high optical depth regime most of the
emission is reflected back to the star.  In fact, resonant up-scattering 
leads to an increase of the radiation temperature by a factor $\sim
2 \beta_{th} \sim 10-20\%$ at most, contrary to what was proposed by
\cite{rude03}.  

On the other hand, estimates of the effective emitting
area based on the observed luminosity and the effective temperature 
 do change due to resonant scattering. 
Radii of neutron stars
are conventionally found from the luminosity and observed temperature,
$R_{bb} \sim ( L_{X}/ 4 \pi \sigma_{\mathrm{SB}} T^4)^{1/2}$,
 where
$L_{X}$ is the observed 
X-ray luminosity, $\sigma_{\mathrm{SB}}$ is
Stefan-Boltzmann constant and $T$ is  the temperature inferred from the spectral fit.
This is called blackbody radius.
Resonant
scatering changes estimates of both  temperature   and luminosity and thus
changes estimate of  neutron star radius.
If
effects of resonant Comptonization are important, then the surface
temperature is lower than the observed one: $T_s \sim T_X /(1+2 a
\beta_{th})$, 
(where $0<a<1$ is a coefficient dependent on the resonant optical depth).
On the other hand, the observed luminosity has contributions both from
the thermal surface emission and from the energy of the scattering
electrons. In case the number of photons is conserved
 the total luminosity is $L_{X} \sim (1+2 a
\beta_{th}) L_{X,s}$. Thus, {\it  the real radius  is larger than
blackbody radius:}
\be 
R_{NS} \sim ( L_{X,s} / 4 \pi \sigma_{SB} T_s^4)
^{1/2} = R_{bb}
  (1+2 a
\beta_{th})^{3/2}
\ee
This may have important implications for attempts to determine
equation of state of neutron stars.

\subsection{Accreting neutron stars}

For sources accreting through the magnetosphere at a residual rate
 $\dot{M}$, the optical depth at energy $\epsilon $ is
\begin{equation}
\tau_{res}\sim { \pi  e \dot{M} \over \sqrt{ r_{res} G M} m_p B_{res} }
\sim 1 \left( {\dot{M} \over 10^{-13} M_\odot/yr} \right)
b^{-7/6}  \left({ \epsilon \over  \mathrm{ 1 keV} } \right) ^{17/6}
\label{dotM}
\end{equation}
with $r_{res}$ given by Eq. (\ref{rres}).
This accretion rate is larger than the Bondi rate for a neutron star
moving at $\sim 100$ km/sec through the ISM, $\dot{M}_B \sim 10^{-17}
M_\odot$/yr, so that isolated neutron stars which are not high energy
pulsars or magnetars cannot support dense, hot magnetospheres (cold
plasma may be supported by radiation pressure \cite{derm91}).  On the
other hand, the accretion rate (Eq.~\ref{dotM}) is considerably
smaller than is typically reached in X-ray binaries.  It is expected
that most of the accreting material is either channeled along the
polar \Bf lines or is flung out by the propeller effect, but even if a
small amount of accreting material gets into the bulk of the neutron star's
magnetosphere (\eg due to development of resistive instabilities in
the boundary layer) the resulting resonant cyclotron optical depth can
be considerable.

In addition, even in the accretion column, the effects of \Bf
inhomogeneities are non-negligible since the size of the accretion
column $H$, which is typically tens of meters, may become comparable
to the width of the resonance layer, which for temperatures of $\sim 5$
keV  ($\beta_{th} =0.01$) and dipole magnetic field  $L_B \sim
r_{NS}/3$ is $\delta r = \beta_{th} L_B \sim 30$ meters.  Thus, effects
of \Bf inhomogeneities should introduce corrections of the order of
unity to resonant radiative transfer in the accretion column.

\section{Discussion}

An astrophysically small amount of matter, heated by some external source
of energy, may considerably modify the emergent spectra through
resonant cyclotron scattering in the magnetospheres of neutron stars and thus
impede the effort to understand the structure and composition of neutron stars
and their equation of state.  The way emission is modified is
qualitatively different both from the case of non-resonant Thomson
scattering as well as resonant scattering in homogeneous \Bfs.

During resonant Comptonization the number of photons is conserved,
while their energy increases by a factor of $\sim 1+2 \beta_{th}$. This
photon energy increase comes at the expense of the electron
energy. Since the minimal charge densities required to produce an
optical depth of the order of unity are small (Eq.~\ref{N}), leptons
are efficiently cooled by Comptonization, so that to be in steady
state they should be heated by some energy source. The power required
is $\sim \beta_{th} L_X$ where $L_X$ is the observed X-ray luminosity.
Depending on the nature of an object heating can be achieved by
different sources.  In young pulsars this may be achieved by pair
production by high energy $\gamma$-ray emission \citep{wrhz98}.  In
magnetars, charges are pulled from the surface by inductive electric
fields generated due to slow crustal motion \citep{tlk02}.  In X-ray
binaries residual accretion through the magnetosphere will heat the
plasma due to the release of gravitational energy.

Resonant Comptonization by mildly relativistic plasma generically
results in soft non-thermal spectra, similar to what is observed in
anomalous X-ray pulsars at intermediate energies, $\leq 10 $ keV.
Non-thermal effects are most prominent for mildly relativistic plasma
$\beta_{th} \leq 1$ at intermediate optical depths, $\tau_{res}\sim 1$.  On
the other hand, at large optical depths, $\tau_{res}\gg 1$ transmitted
spectrum  is nearly thermal, but shifted to higher frequencies.

An important advantage of the model 
is that in our case photons are
just moved to higher energies, so that one cannot simply extrapolate
the data from the X-rays down to longer wavelengths. In the
conventional blackbody plus power law fitting procedure one also needs
to assume a low energy cut-off to the power law,
\cite[\cf][]{halpgott05}.  Such a cutoff is not required in our model.

As a principal application of the model, we have fitted the  non-thermal
spectrum of AXP \tfe\ and found a quantitatively good agreement.
Besides the production of non-thermal spectra in magnetars, resonant
scattering may change estimates of the surface temperature while
conserving the overall Plankian form of the spectrum (in case of high
optical depths), and may explain the lack of surface spectral lines.
Variations of magnetospheric plasma parameters may induce spectral
changes on time scales much shorter than the evolutionary time scale
of neutron stars.

In this paper we addressed the principal issues related to resonant
cyclotron transfer, employing a simple one-dimensional model of
resonant radiation transfer in the limit of non-relativistic
temperatures of scattering particles with a water-bag distribution.
More involved three-dimensional Monte-Carlo simulations are in
progress. They will extend the current approach to 
polarized surface emission.   The magnetospheric plasma may be
relativistically hot, $\beta_{th} \sim 1$, in which case the diffusion of
photons both in frequency and in real space is nonlocal and, in
addition, scattering on high harmonics becomes important.  Also,
possible bulk motion of the scattering plasma may lead to the additional
effect of bulk Comptonization. Resonant scattering by relativistic
electrons has been invoked by \cite{thompbelo04} to explain the high
energy emission from magnetars.  Finally, the fact that surface
radiation is not blackbody \citep[\eg][]{zavp02} should also be
incorporated.  We plan to address some of these issues in a
subsequent publication.

\acknowledgements We would like to thank Victoria Kaspi and
Christopher Thompson for
numerous discussion and insightful comments. We are grateful to Peter
M. Woods for his help reducing the \cxo\ observation of \tfe. We also
would like to thank W. Ho, K. Mori, M. Ruderman, M. van Kerkwijk  for
comments and discussions.

\appendix

\section{One-dimensional Monte-Carlo simulation}
\label{MC}

\begin{figure}
\plotone{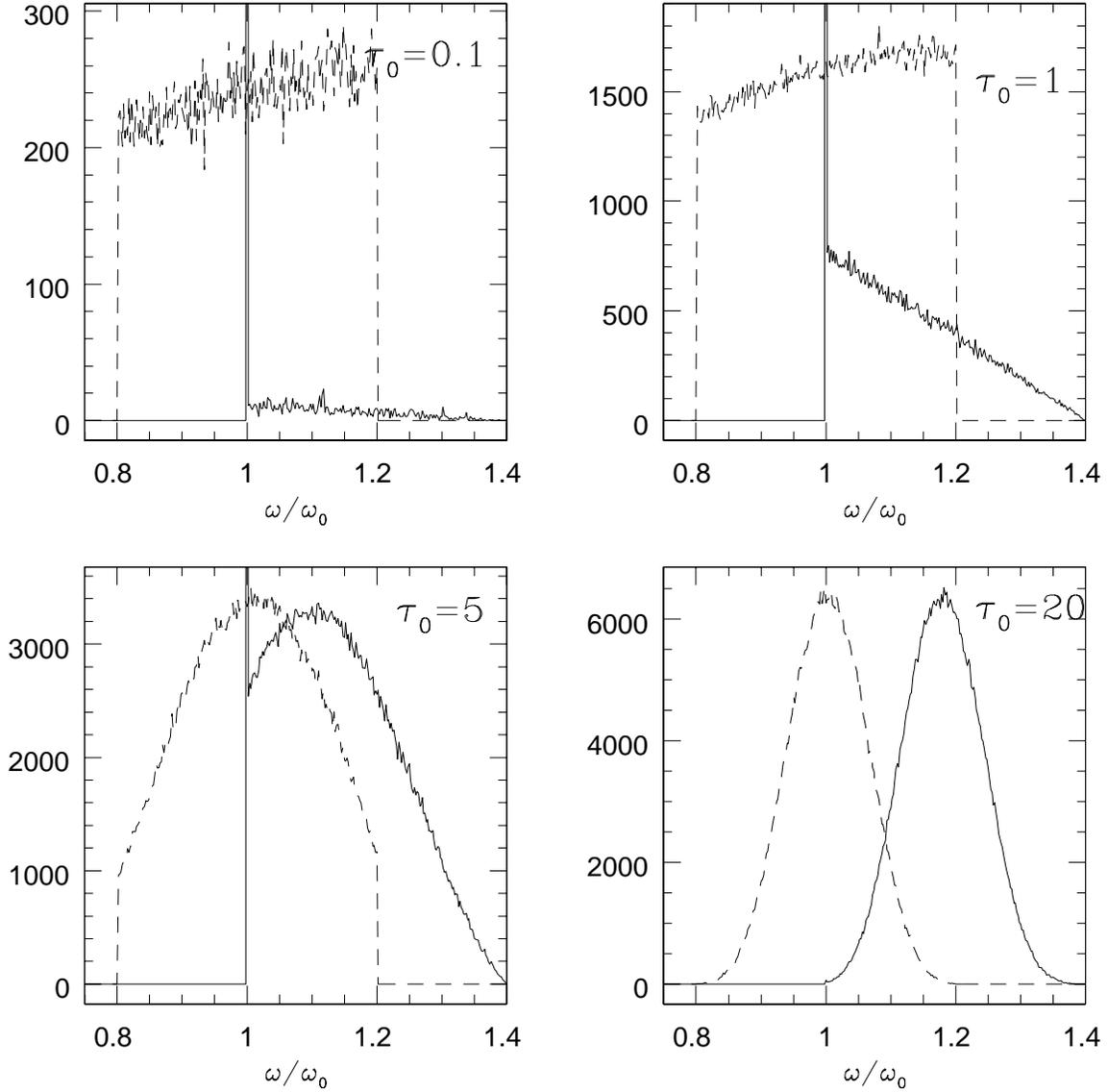}
\caption{Same as Fig.~\ref{resscat} reproduced using one-dimensional
Monte-Carlo simulations with $10^6$ photons and taking the  $1-\beta$
enhancement of the cross-section into account, $\beta_{th} = 0.1$.  At
intermediate optical depths the reflected flux is skewed towards higher
frequencies due to the kinematic enhancement of the cross-section.  These
simulations are done using a one-dimension approximation for constant
$L_B$.  }
\label{resscatMC1}
\end{figure}

Analytical results in Section \ref{sec:main} have been obtained in the
limit $\beta \rightarrow 0$, neglecting the kinematic enhancement of
the scattering rate by a factor $1-\beta$ due to the motion of a
particle. To test the validity of this approximation we perform
one-dimensional Monte Carlo simulations of the transfer of resonant
cyclotron radiation.  In the limit $\beta \ll 1$ we can expand
frequencies about the initial frequency $(\om-\om_0)/\om_0 \ll 1$ and
assume $\om_B = \om_0(1-x)$ where $x$ is a dimensionless coordinate
(measured in terms of $L_B$).  Under these circumstances the resonant
velocity is $\beta= \pm (x + \om - \om_0)$ and at each scattering the
new frequency becomes $ \tilde{\om } = 2 \om_0(1-x) - \om$.  At each
step the probability of scattering backwards is multiplied by a
kinematic factor $1-\beta$.  Results for the transmission of a
spectral line are present in Fig.~\ref{resscatMC1}.  At intermediate
optical depths back scattered photons have on average higher
frequencies. This is due to the fact that photons are more likely to
encounter backward propagating electrons, whose cross-section is
enhanced by $1+\beta$.  On the other hand, at high optical depth
deviations from analytical results are small. In this limit photons
experience multiple scatterings inside the layer so that the relative
velocity of the electrons between each scatter is small and the
kinematic enhancement is not important.

\begin{deluxetable}{cccc}
\tablecolumns{2}
\tablewidth{\columnwidth}
\tablecaption{Phase-averaged spectral fits of \tfe\ for various models.\label{ta:spectra}} 
\tablecolumns{4}
\tablehead{\colhead{Parameter}&  \colhead{Value\tablenotemark{a}}  }
\startdata
\cutinhead{Resonant Compton Scattering Model}
$N_{H}$ (10$^{22}$ cm$^{-2}$)  & $0.76 \pm 0.16$  \\
$k_BT$ (keV)		       & $0.39 \pm 0.05 $  \\ 
$\beta$   		       & 0.3 (fixed)      \\
$\tau_{res}$                         & 3 (fixed)       \\
Flux\tablenotemark{b} (10$^{-12}$ ergs cm$^{-2}$ s$^{-1}$)  & 7.61 \\
Unabs Flux\tablenotemark{b} (10$^{-12}$ ergs cm$^{-2}$ s$^{-1}$) & 8.64 \\
$\chi^2$/dof                   &  293/278    \\
\cutinhead{Blackbody + Power Law}   
$N_{H}$ (10$^{22}$ cm$^{-2}$)  & $1.18 \pm 0.04$   \\
$k_BT$ (keV)		       & $0.585\pm 0.012$  \\ 
$\Gamma$   		       & $3.08 \pm 0.10$   \\
Flux\tablenotemark{b} (10$^{-12}$ ergs cm$^{-2}$ s$^{-1}$)  & $7.91$  \\
Unabs Flux\tablenotemark{b} (10$^{-12}$ ergs cm$^{-2}$ s$^{-1}$) & 9.56 \\
$\chi^2$/dof                   &  274/276    \\
\cutinhead{Blackbody + Blackbody}   
$N_{H}$ (10$^{22}$ cm$^{-2}$)  & $0.76 \pm 0.03$   \\
$k_BT_1$ (keV)		       & $0.458 \pm 0.030$  \\ 
$k_BT_2$ (keV)  		       & $0.843 \pm 0.058$   \\
Flux\tablenotemark{b} (10$^{-12}$ ergs cm$^{-2}$ s$^{-1}$)  & $7.61$ \\
Unabs Flux\tablenotemark{b} (10$^{-12}$ ergs cm$^{-2}$ s$^{-1}$) & 8.62 \\
$\chi^2$/dof                   &  284/276    \\

\enddata

\tablenotetext{a}{The quoted uncertainties reflect the 1$\sigma$ error
for a reduced $\chi^2$ of unity.}  \tablenotetext{b}{Flux in the
2$-$10 keV band.}
\end{deluxetable}

\end{document}